\documentclass[a4paper,11pt]{article}

\pdfoutput=1

\usepackage{contribution}

\newcommand{\weblink}[2][]{%
    \ifthenelse{\equal{#1}{}}%
    {\textnormal{\url{#2}}}%
    {\textnormal{\href{#2}{#1}}}%
}

\newcommand{\acknowledgements}[1]{%
  \bigskip\bigskip
  \textsf{\textbf{\Large Acknowledgements}} \\[2ex]
  {#1}
  \bigskip
}

\def\beq{\begin{equation}}
\def\eeq#1{\label{#1}\end{equation}}
\def\eeqn{\end{equation}}

\def\beqa{\begin{eqnarray}}
\def\eeqa#1{\label{#1}\end{eqnarray}}
\def\eeqan{\end{eqnarray}}

\let\bar=\overbar

\def\etal{{\it et al.}}
\def\ie{{\it i.e.}}

\def\Dslash{\not{\hbox{\kern-4pt $D$}}}
\def\dslash{\not{\hbox{\kern-2pt $\del$}}}

\def\msb{{\bar{\ssstyle M \kern -1pt S}}}

\newcommand{\contribution}[7][]{%
  \clearpage
  \thispagestyle{plain}
  \ifthenelse{\equal{#1}{}}
  {\hypersetup{pdftitle={#2}}}
  {\hypersetup{pdftitle={#1}}}
  \hypersetup{pdfauthor={{#3} {#4}}}
  {\centering\normalfont\LARGE\bfseries\sffamily #2 \par\nobreak}
  \lhead{}
  \chead{%
    \textit{\footnotesize XIV International Conference on Hadron Spectroscopy
      (\weblink[\textit{hadron2011}]{http://www.hadron2011.de}), 13-17 June 2011, Munich, Germany}%
  }
  \rhead{}
  \bigskip
  \begin{center}
    {#3} {#4}\ifthenelse{\equal{#6}{}}{}{\footnote{\weblink[#6]{mailto:#6}}}
    \ifthenelse{\equal{#7}{}}{}{#7} \\
    \textit{#5}
  \end{center}
  \bigskip
}

\renewcommand{\abstract}[1]{%
  \begin{center}
    \begin{minipage}{0.85\textwidth}
      \begin{footnotesize}
        #1
      \end{footnotesize}
    \end{minipage}
  \end{center}
  \bigskip
}

\begin{document}

%
%
%
%
%
{  

\makeatletter
\@ifundefined{c@affiliation}%
{\newcounter{affiliation}}{}%
\makeatother
\newcommand{\affiliation}[2][]{\setcounter{affiliation}{#2}%
  \ensuremath{{^{\alph{affiliation}}}\text{#1}}}
%

\contribution[$\pi^{-}\gamma \rightarrow \pi^{-}\pi^{-}\pi^{+}$ at Low Masses compared to ChPT Prediction]
{Measurement of $\mathbold{\pi^{-}\gamma \rightarrow \pi^{-}\pi^{-}\pi^{+}}$ at Low Masses, and Comparison to ChPT Prediction, at COMPASS}
{Stefanie}{Grabm\"uller}  
{\affiliation[Technische Universit\"at M\"unchen, Physik-Department, 85748 Garching, GERMANY]{1} \\
 \affiliation[State Research Center of the Russian Federation, IHEP, 142281 Protvino, RUSSIA]{2}} 
{stefanie.grabmueller@tum.de}
{\!\!$^,\affiliation{1}$, Dmitry Ryabchikov\affiliation{2}, and Jan M.~Friedrich\affiliation{1} \newline on behalf of the COMPASS Collaboration}

%


%

\abstract{%
This paper presents an analysis of $\pi^{-}{\rm Pb} \rightarrow X^{-}{\rm Pb} \rightarrow \pi^{-}\pi^{-}\pi^{+} {\rm Pb}$ 
events at $190\,{\rm GeV}/c$ beam momentum and very low four-momentum transfer $t'<0.001\,{\rm GeV}^{2}/c^{2}$. 
Coherent scattering off the nucleus as a whole dominates with contributions from Reggeon, Pomeron and photon exchange.
The latter originates from Primakoff reactions and is identified by the sharp Coulomb peak of intensities at $t' \approx 0$.
The partial-wave analysis of these data focusses on new techniques for the extraction of the Primakoff contribution at low masses. 
Its measured absolute cross-section at $\sqrt{s} < 5 m_{\pi}$ is well in agreement with the prediction from chiral perturbation theory.

}
%

\section{Meson Spectroscopy at Low Momentum Transfer}

Dissociation of pions on nuclear or hydrogen targets provides clean access to the light meson spectrum.
In case of heavy target nuclei and low momentum transfer, mesons can be produced by two interaction mechanisms:
diffractive production by Reggeon $t'$-channel exchange and Primakoff (or Coulomb) production via
the exchange of quasi-real photons. The latter has its dominant contribution at vanishing momentum transfer and 
introduces spin-projection $M = \pm 1$ to the produced system.
At low masses, \ie at the threshold of the investigated final state, the only involved hadrons are pions scattering with each other and the photon.
Thus predictions from chiral perturbation theory (ChPT), which take into account coupling to a real photon simultaneously, are supposed to be applicable.
At higher masses, resonances can be produced, so that the radiative coupling of the $a_{2}(1320)$, and potentially also of heavier mesons, 
can be investigated. Also the interference between the Coulomb and diffractive production of these resonances can be studied.

COMPASS is a multi-purpose fixed-target experiment at the CERN SPS, that investigates the structure and spectroscopy of hadrons.
The two-stage high-precision spectrometer \cite{compass_spectro} can detect outgoing particles
 within a large range of scattering angles and particle momenta,
and provides a uniform acceptance, especially for reactions featuring low to intermediate momentum transfer from the beam to the target.
For a short run with a $190\,{\rm GeV}/c$ hadron beam (composed of $96.8\,\%\,\pi^{-}$, $2.4\,\%\,K^{-}$ and $0.8\,\%\,\bar{p}$)
 on thin lead disk targets in 2004, additional silicon micro-strip detectors were installed downstream of the target
 to resolve also smallest scattering angles. 
With this setup, about 4 million exclusive $\pi^{-}\pi^{-}\pi^{+}$ events have been collected
 with a dedicated multiplicity trigger for charged-particle final states. 
About 1 million of these feature very low momentum transfer $t'<0.001\,{\rm GeV}^{2}/c^{2}$, 
with $t' = |t|-|t|_{{\rm min}}$ constituted of the squared four-momentum transfer $t$ from the beam
to the produced system, and $|t|_{{\rm min}}$ the minimum value of $|t|$ allowed by kinematics. 
In this region, referred to as "Primakoff $t'$", photon exchange appears to be relevant in the data.

Partial-wave analysis (PWA) strives for determining all resonances present in a given data set and their properties 
by fitting angular distributions and taking into account interference effects.
The PWA presented here is based on the commonly-known isobar model, that was adapted to the specific needs of the analysis.
According to this model, the produced resonances decay via intermediate two-particle decays into the particles
observed in the spectrometer. 
For $3\pi$ events, a partial wave in the reflectivity basis is written as $J^{PC}M^{\epsilon}[isobar\;\pi]L$, defining the quantum numbers of the resonance, $J^{PC}$, spin projection $M$, reflectivity $\epsilon$, the isobar, and the angular momentum $L$ between the di-pion resonance and the unpaired pion. The PWA method and its basic assumptions are described in more detail in Ref.~\cite{compass_exotic} and the references therein.

\section{Diagrams from ChPT}

\begin{figure}[b]
  \begin{center}
    \includegraphics[width=0.32\textwidth]{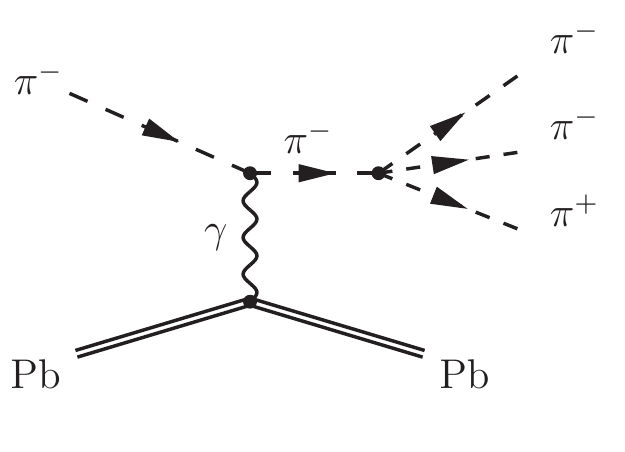}
    \includegraphics[width=0.32\textwidth]{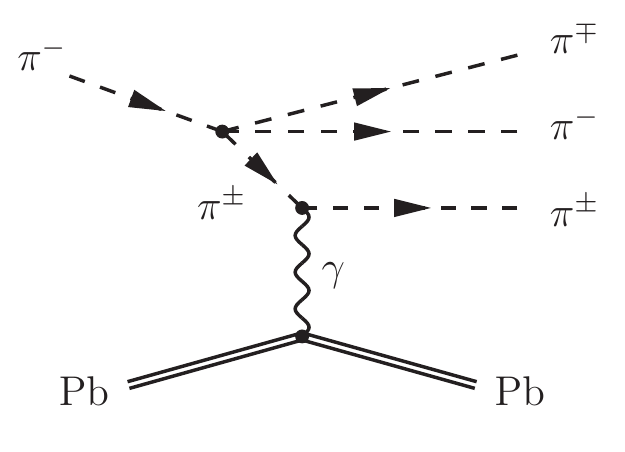}
    \includegraphics[width=0.32\textwidth]{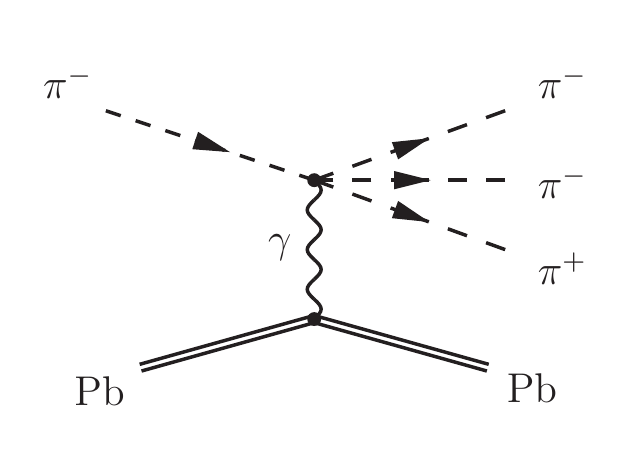}
    \caption{Leading order processes in ChPT \cite{KaiserFriedrich08} for the reaction $\pi^{-}\gamma \rightarrow \pi^{-}\pi^{-}\pi^{+}$, embedded in the Primakoff reaction contribution to $\pi^{-}{\rm Pb} \rightarrow \pi^{-}\pi^{-}\pi^{+} {\rm Pb}$.}
    \label{fig:chpt_tree}
  \end{center}
\end{figure}

In the Primakoff $t'$ region, decay amplitudes with $M=1$ can be attributed to Primakoff production,
\ie effectively $\pi^{-}\gamma \rightarrow X^{-} \rightarrow \pi^{-}\pi^{-}\pi^{+}$ scattering when observing 
$\pi^{-}{\rm Pb} \rightarrow X^{-}{\rm Pb} \rightarrow \pi^{-}\pi^{-}\pi^{+} {\rm Pb}$. 
In addition the low mass region, i.e. $m_{3\pi} < 0.72\,{\rm GeV}/c^{2}$, is governed by pion-scattering
instead of resonance production so that ChPT calculations are applicable.
The leading order processes in ChPT are calculated in \cite{KaiserFriedrich08} and depicted in Fig.~\ref{fig:chpt_tree}, including also the lead nucleus which is not part of the calculation.
As it has been shown in \cite{Kaiser10} that for low masses the next order in ChPT (i.e. loops) does not change the cross-section of $\pi\gamma$-scattering, $\sigma_{\gamma}$, appreciably,
 it is sufficient to introduce the leading order calculations for $\sigma_{\gamma}$ into the PWA,
multiplied with the quasi-real photon density provided by a nucleus of charge $Z$ as given by Weizs\"acker and Williams \cite{WW61}.
The experimentally observed cross-section is given by
\begin{equation}
  \label{eq:WW-cross-sec}
  \frac{d\sigma_{\rm Pb}}{ds\, dt'\, d\Phi_{n}}\ =\ 
  \frac{\alpha\cdot Z^2}{\pi (s-m_\pi^2)}\cdot
  \frac{t' \cdot F_{\mbox{\tiny eff}}^2(t') }{(t'+t_{\min})^2} \cdot
  \frac{d \sigma_{\gamma}}{d\Phi_{n}} 
\end{equation}
with $d\Phi_{n}$ the phase-space element for the final state system $X^{-}$ and $ F_{\mbox{\tiny eff}}^2(t')$
the effective lead form factor,
and is introduced in its fully differential form as an additional decay amplitude with $M=1$ to the PWA fit. 
This special partial wave, called "chiral amplitude", is mathematically not orthogonal 
to the $M=1$ waves of the isobar model, so that these have in fact to be replaced by the chiral amplitude in the low mass region.
The resulting fit quality is compatible with the quality of a previous fit that used up to 6 waves with the $(\pi\pi)_{s}$ wave or the $\rho(770)$ as isobars,
but the chiral-amplitude fit works with much less fitting parameters.
At masses above $m_{3\pi}>0.72\,{\rm GeV}/c^{2}$ isobaric decays that compete with the chiral amplitude have to be taken into account again.

\begin{figure}[tb]
  \begin{center}
    \includegraphics[width=0.48\textwidth]{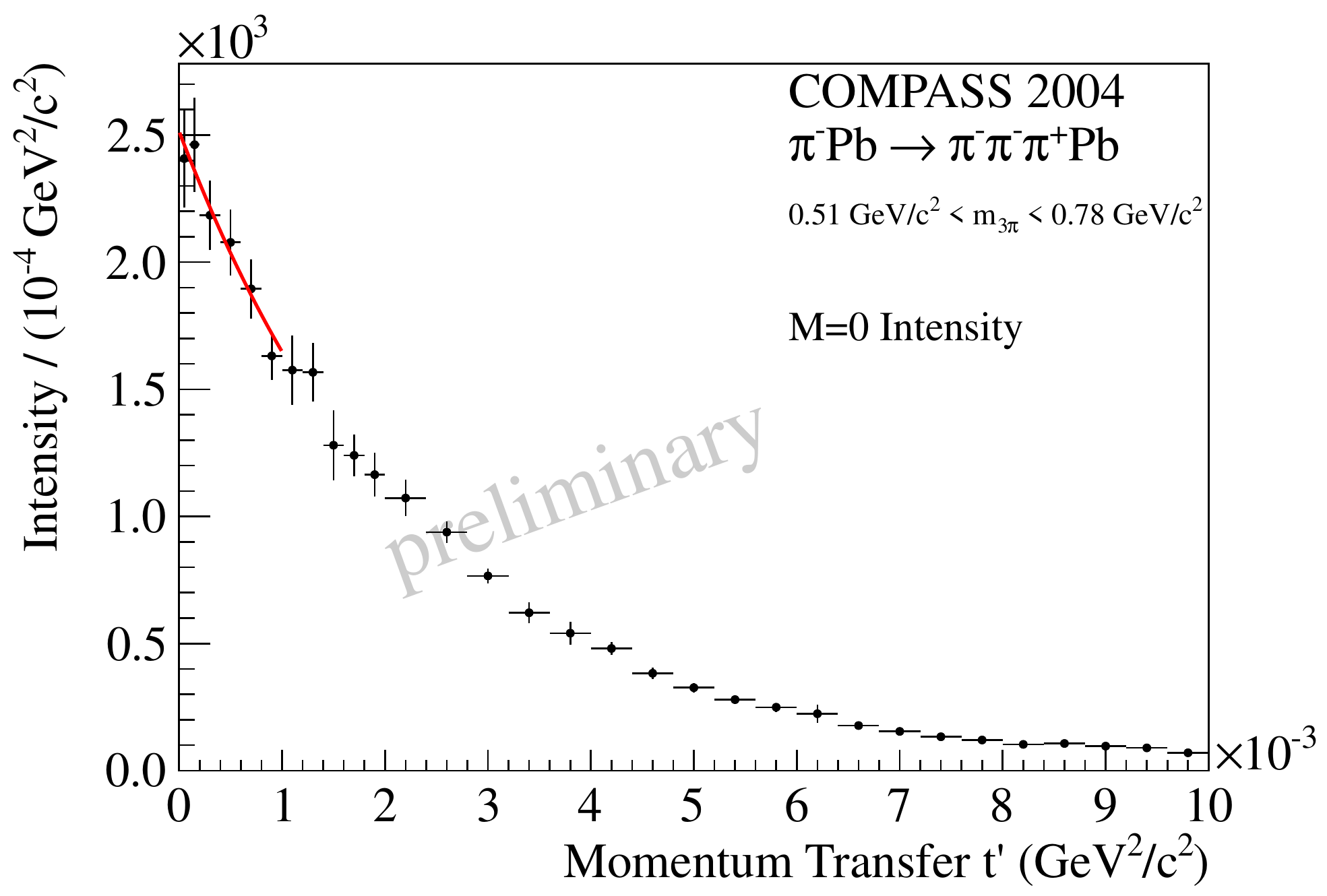}
    \includegraphics[width=0.48\textwidth]{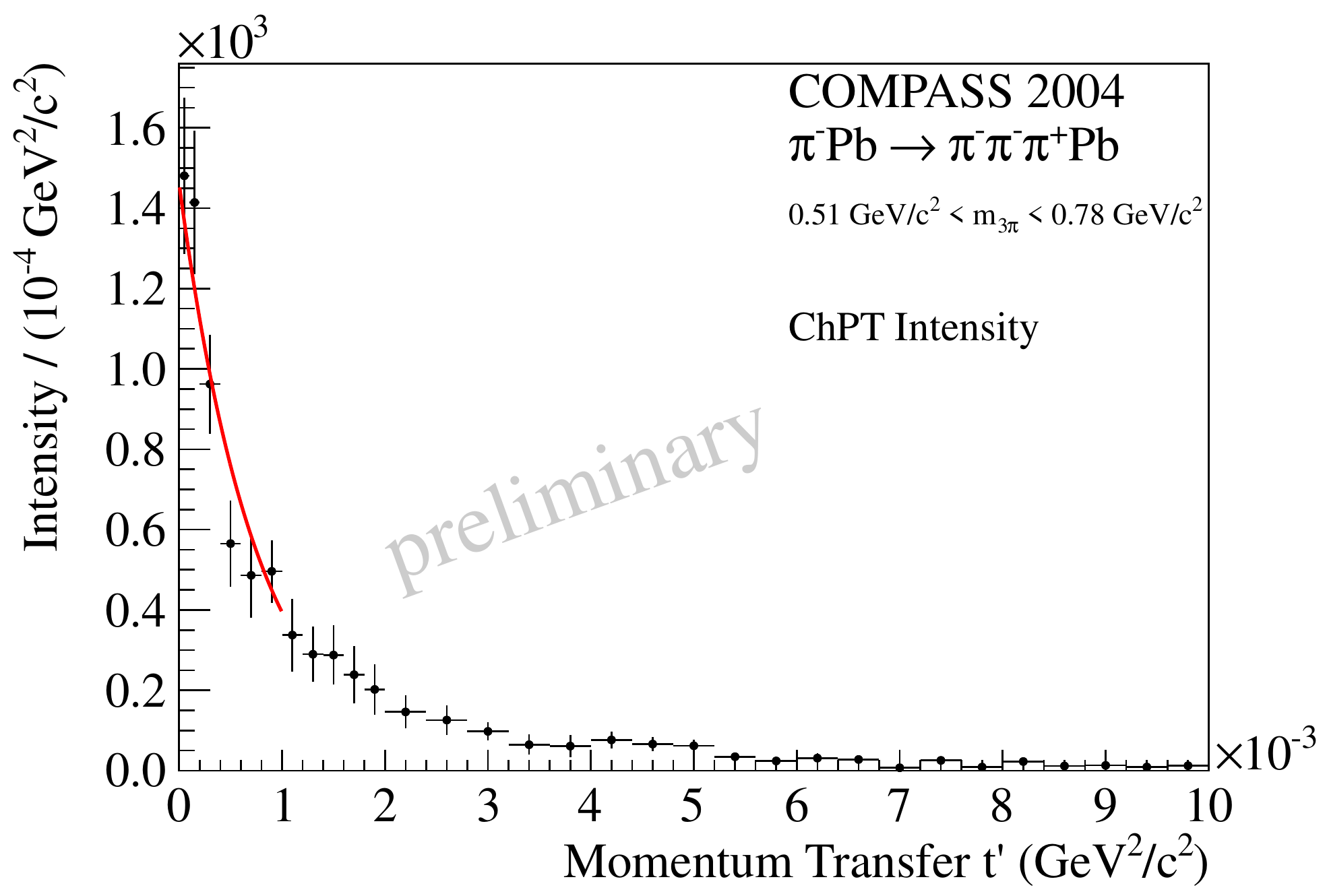}
    \caption{$t'$ dependences of the $M=0$ total intensity (left) and the intensity of the chiral amplitude (right) confirm the diffractive production of the $M=0$ states as well as the Primakoff nature of the intensity picked up by the chiral amplitude.}
    \label{fig:tdist}
  \end{center}
\end{figure}

In order to get a confirmation of the production mechanism that is attributed to the chiral amplitude,
a dedicated set of PWA fits was carried out in several bins of $t'$ in the low-mass region 
without taking into account any $t'$ dependence of the decay amplitudes.
The resulting intensity spectra of the total intensity of the chiral amplitude and the 
 total $M=0$ intensity 
 in this mass region, dependent on $t'$, are shown in Fig.~\ref{fig:tdist}.
Both spectra are fitted by an exponential $\propto \exp(-b t')$ for $t'<0.001\,{\rm GeV}^{2}/c^{2}$.
The $M=0$ intensity features a slope $b \approx 400\,{\rm GeV}^{-2}/c^{-2}$ which is 
 compatible with diffractive production off lead nuclei.
The ChPT intensity is much steeper with $b \approx 1600\,{\rm GeV}^{-2}/c^{-2}$ which 
is consistent with the expected effective slope of Primakoff production if the $t'$ dependence from 
 Eq.~\ref{eq:WW-cross-sec} is smeared with the experimental resolution.

\section{Absolute Cross-Section and ChPT Prediction}

For the comparison with the ChPT prediction from \cite{KaiserFriedrich08},
PWA fits for $0.44 < m_{3\pi} < 0.72\,{\rm GeV}/c^{2}$ were performed
in $40\,{\rm MeV}/c^{2}$ mass bins for $t'<0.001\,{\rm GeV}^{2}/c^{2}$, that use
only the chiral amplitude associated with Primakoff production as $M=1$ contribution.

\begin{figure}[tb]
  \begin{center}
    \includegraphics[width=0.48\textwidth]{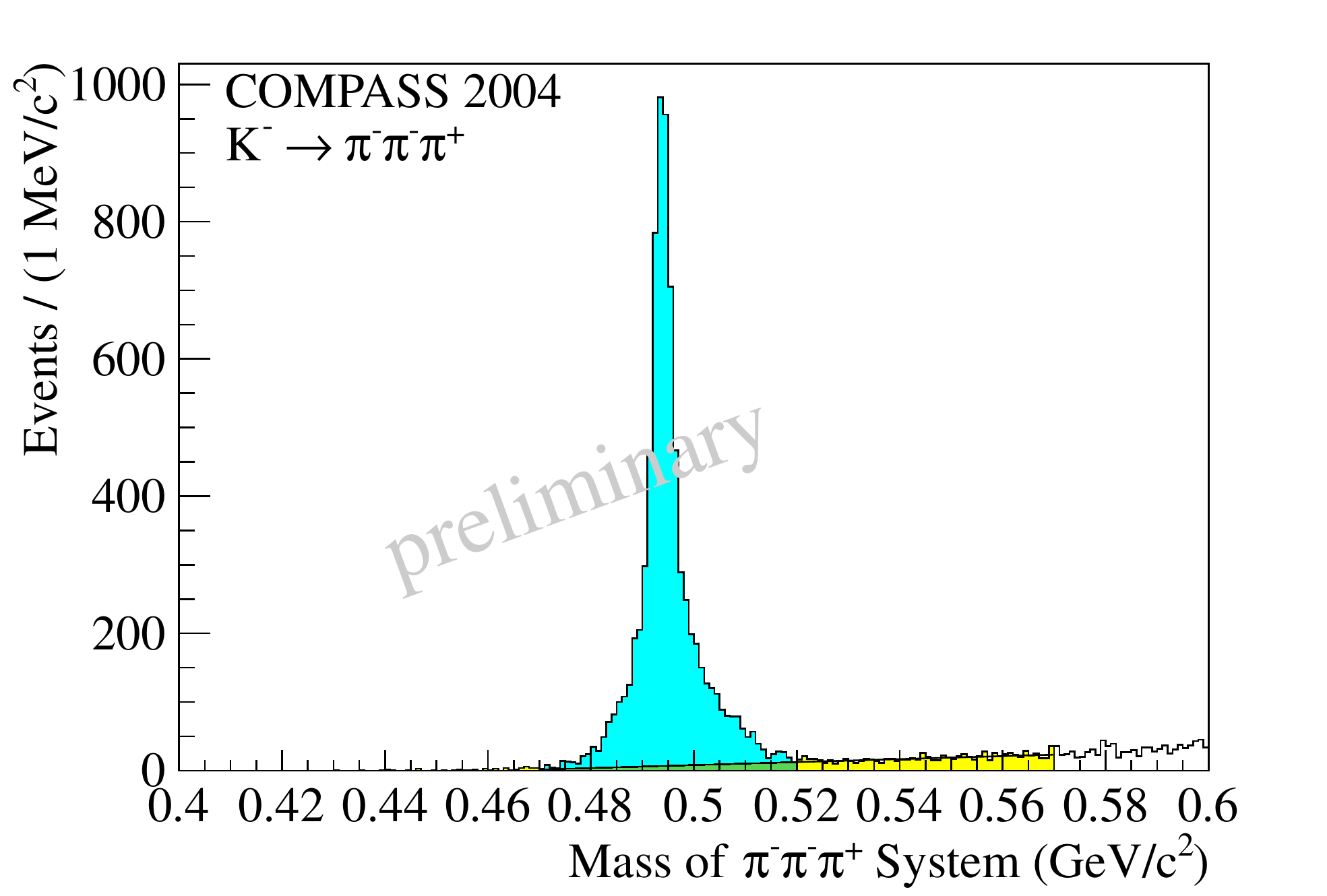}
    \includegraphics[width=0.48\textwidth]{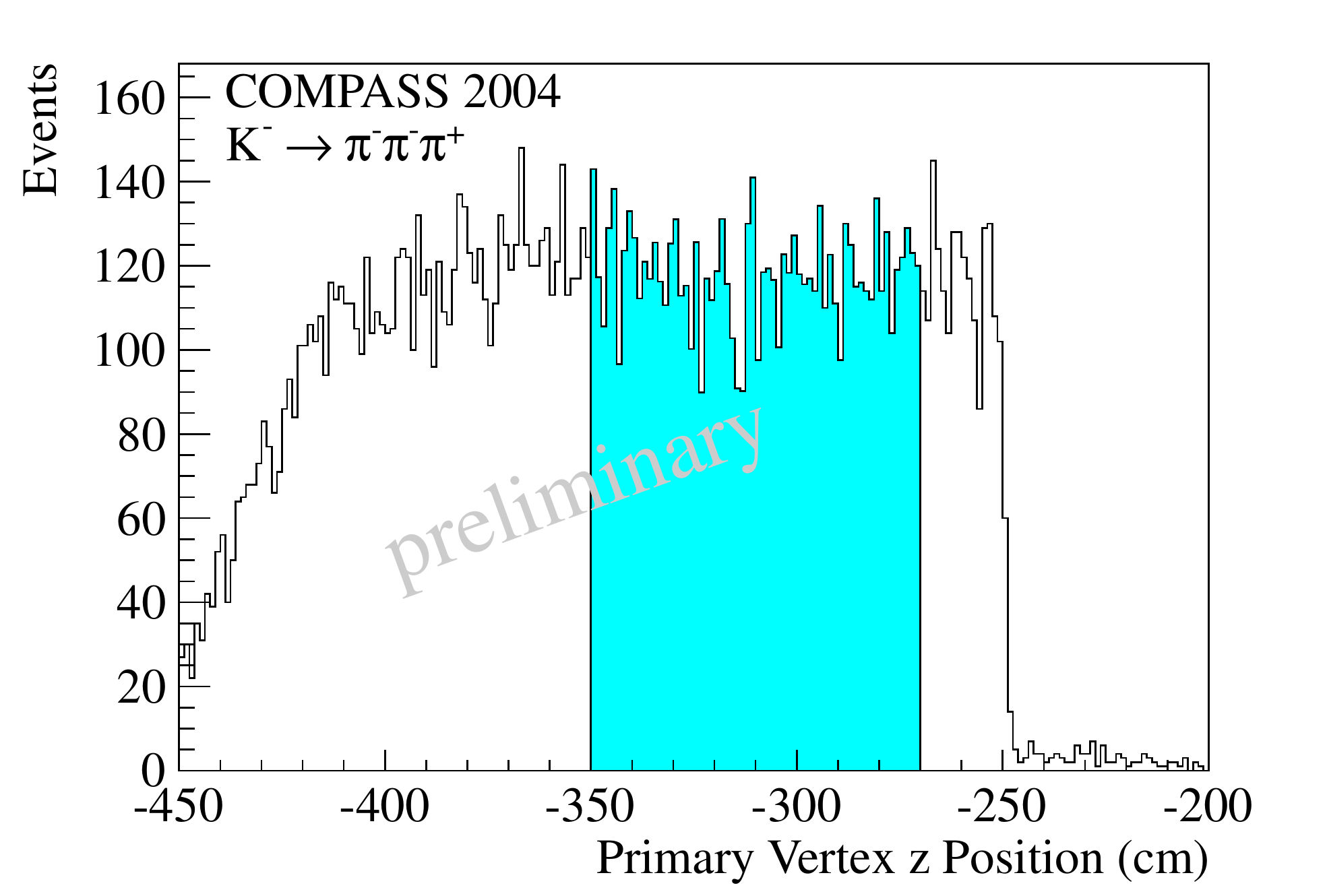}\\
    \includegraphics[width=0.48\textwidth]{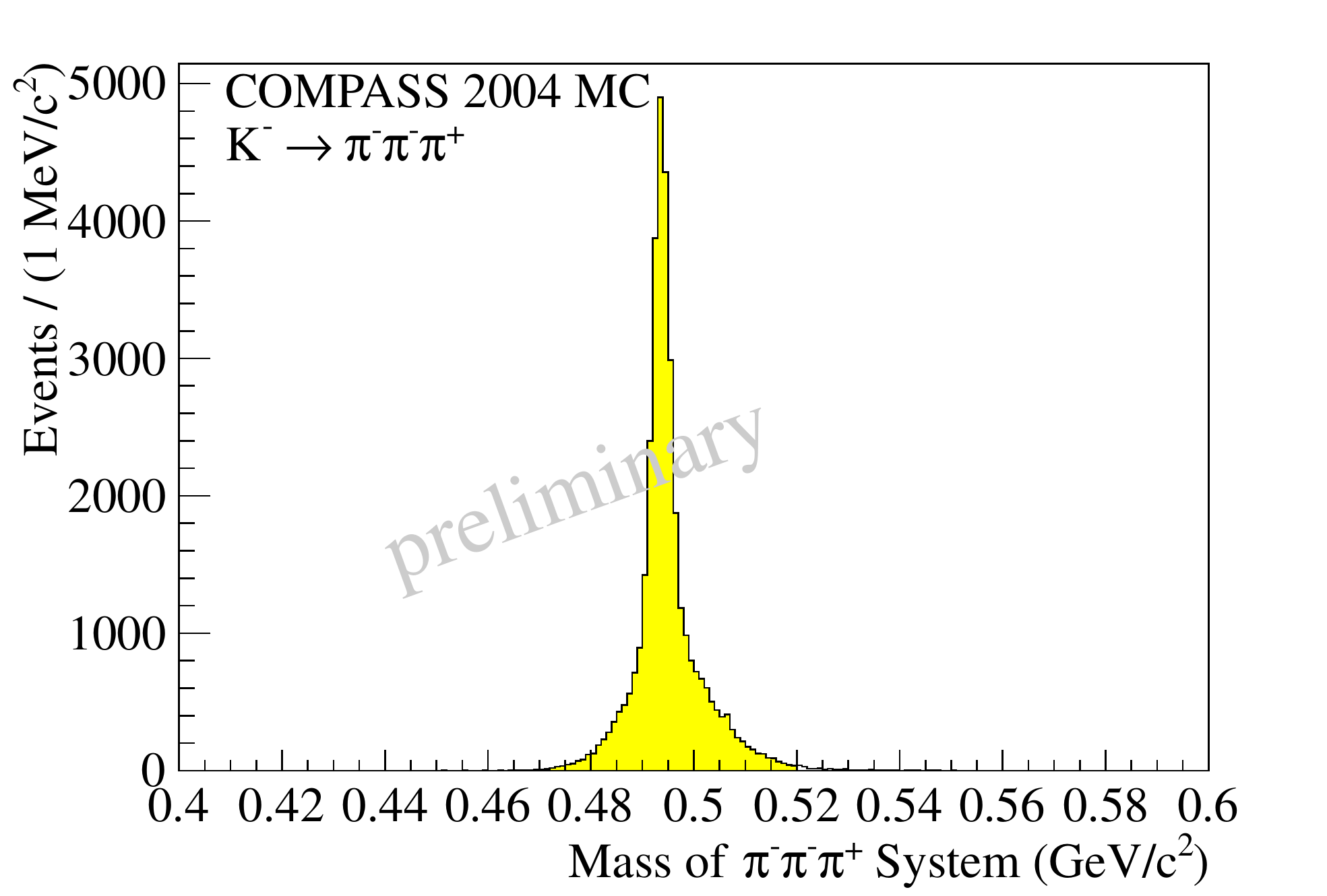}
    \includegraphics[width=0.48\textwidth]{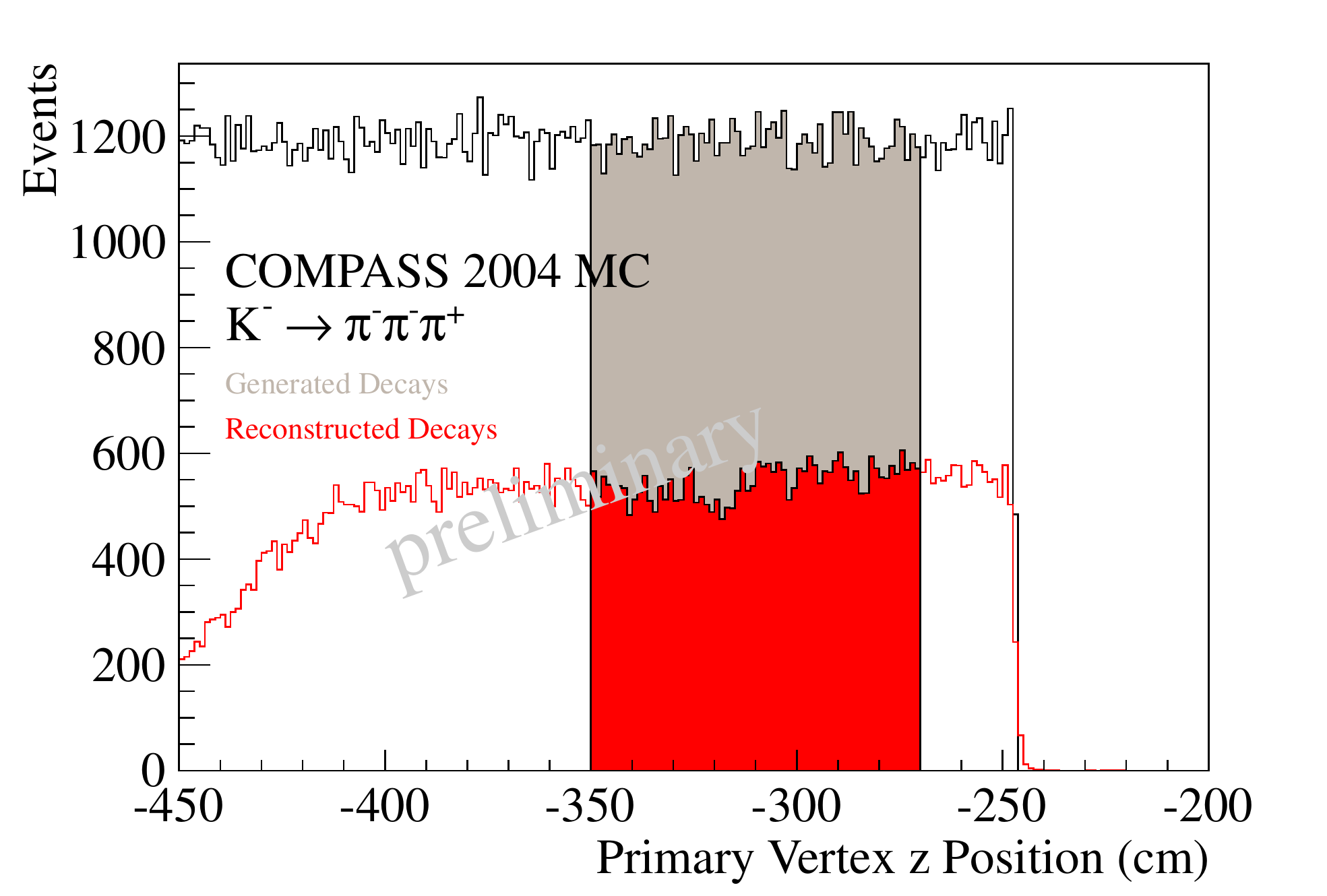}
    \caption{Luminosity determination from the free decay of beam kaons: Reconstructed invariant mass spectra (left) and decay vertex positions (right) for both COMPASS data (top row) and simulated kaon decays in the COMPASS target region (bottom row).}
    \label{fig:kaons}
  \end{center}
\end{figure}

The total intensity of the chiral amplitude, fitted from the experimental data and corrected for acceptance effects,
 is converted into its absolute cross-section via the well-known thickness of the lead target 
 and the incoming beam flux. 
The absolute beam intensity was not monitored reliably, but the effective beam flux can be
 obtained with good precision from the observed $K^{-} \rightarrow \pi^{-}\pi^{-}\pi^{+}$ decays in the COMPASS target region,
 which originate from the kaon component present in the negative hadron beam.
Thus the normalization is available from the same data-taking set 
 (\ie in the same final state, but with broader cut on the decay vertex position)
 featuring the same systematics concerning the relevant trigger and detector efficiencies.
Fig.~\ref{fig:kaons} presents the invariant $3\pi$ mass spectrum, obtained with broad vertex cut, showing the clean kaon signal (top left), 
 from which the observed number of kaons is obtained 
 after subtraction of the small background that is estimated by a linear fit.
The corresponding decay vertex distribution (Fig.~\ref{fig:kaons} top right) shows that
 these decay vertices are reconstructed in free space, 
 at the downstream end of the spectrum sharply limited by the position 
 of the respective charged-particle multiplicity trigger,
 and upstream limited by the positions of the beam telescope detectors defining the vertex reconstruction efficiency.
The background from pion interactions in the lead target is obtained from the vertex distribution in the neighboring mass region,
 scaled accordingly and subtracted statistically. 
The selected region of reconstructed vertex positions, that proceeds to the mass spectrum discussed before, is selected as indicated to assure 
 uniform reconstruction efficiency inside the considered region.
These two distributions were also confirmed by a dedicated full Monte Carlo (MC) simulation of kaon decays
 in the relevant region of the COMPASS spectrometer.
Fig.~\ref{fig:kaons} (bottom left) shows the mass distribution of the reconstructed MC kaon decays. 
The contribution from pion interactions present in the real data as smoothly rising background is now completely absent.
The shape of the reconstructed kaon mass is precisely modeled. 
The broad part at the base of the kaon peak
 is traced back to kaons decaying upstream of the lead target.
In that case all three decay pions undergo multiple scattering in the lead, 
 which leads to the observed distortion.
Fig.~\ref{fig:kaons} (bottom right) presents the distributions for both simulated and reconstructed kaon decay vertices,
 confirming the indicated cuts on the reconstructed vertices in real data to be reasonably chosen.

\begin{figure}[tb]
  \begin{center}
    \includegraphics[width=0.48\textwidth]{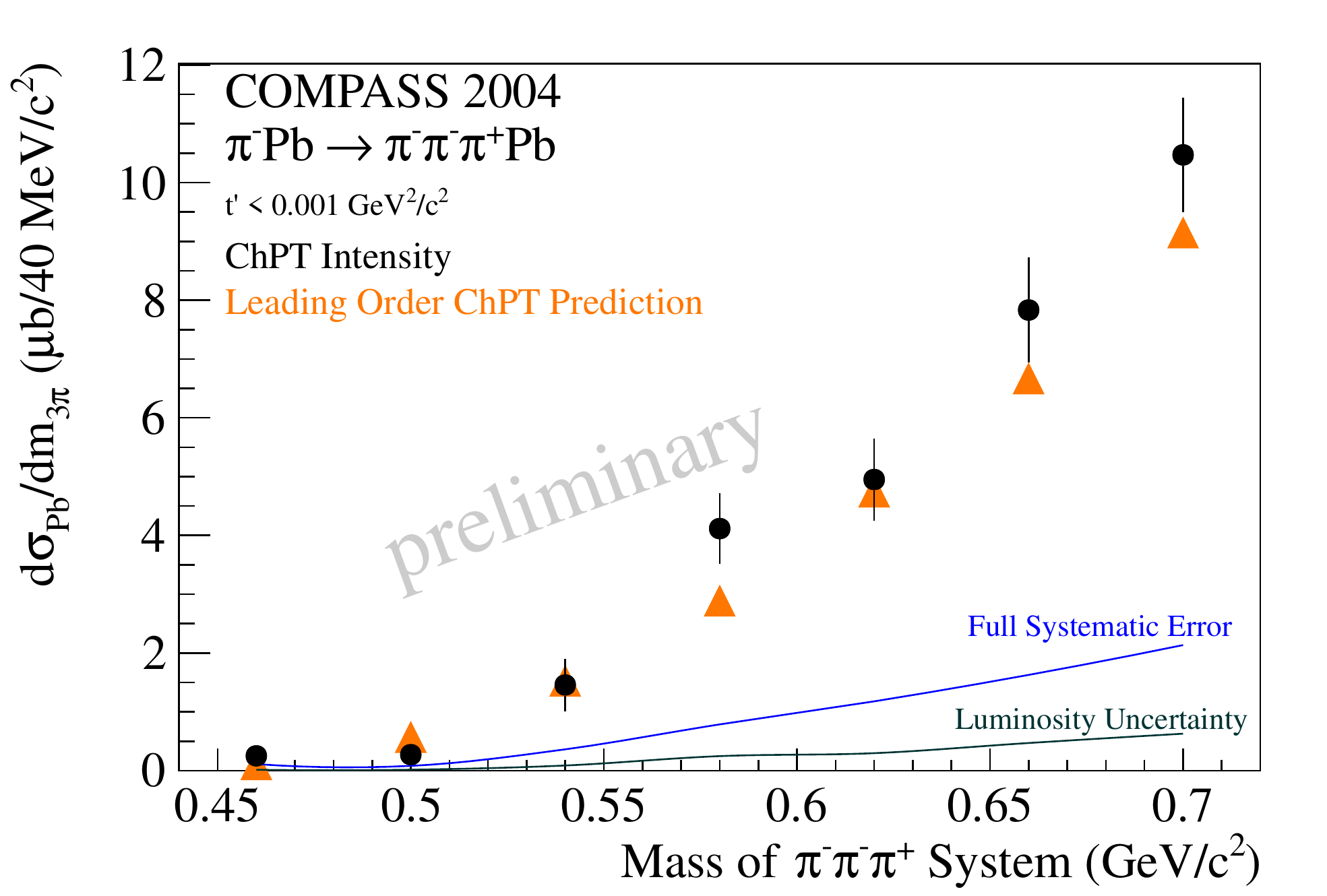}
    \includegraphics[width=0.48\textwidth]{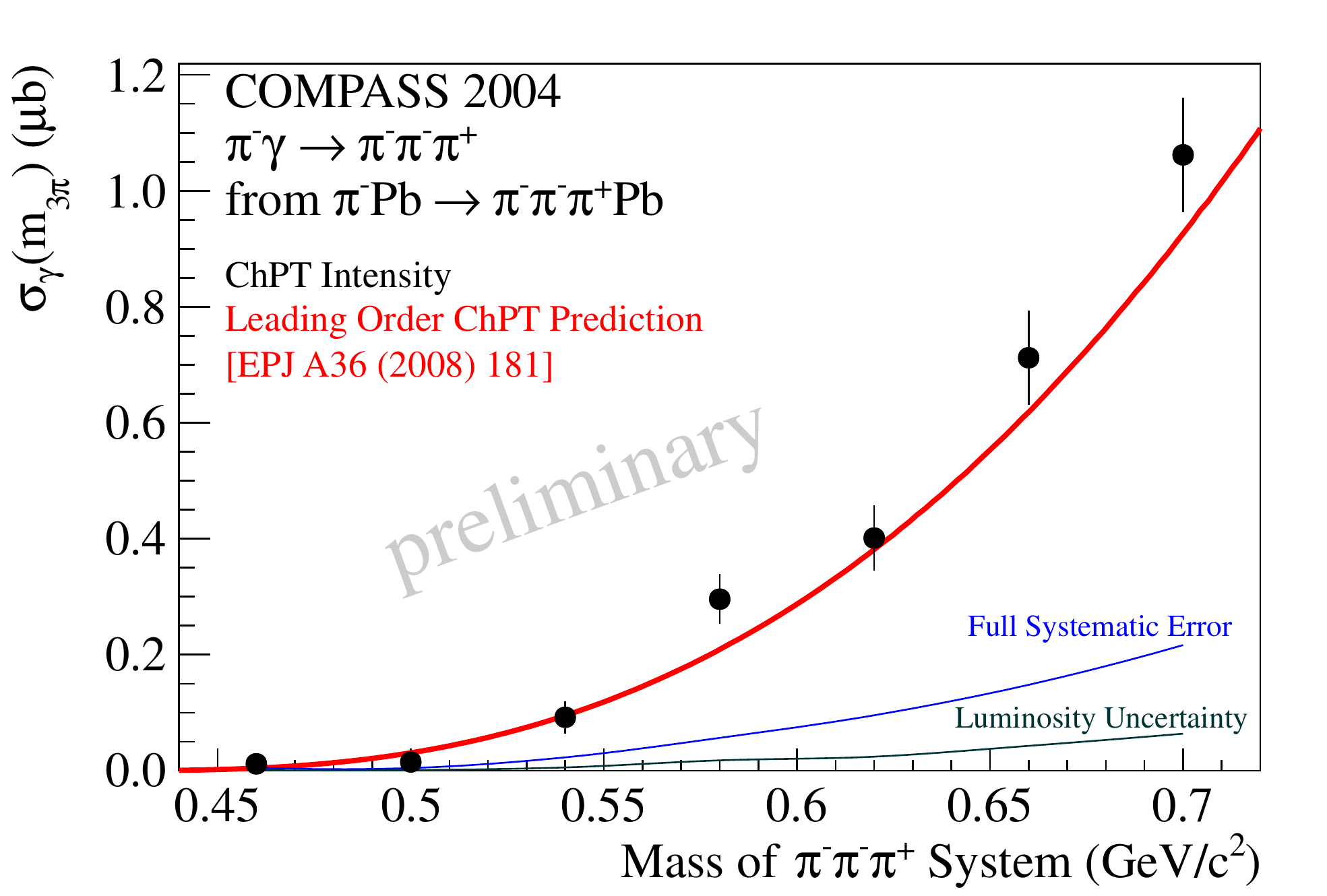}
    \caption{Measured cross-section of $\pi^{-}{\rm Pb} \rightarrow \pi^{-}\pi^{-}\pi^{+} {\rm Pb}$ induced by photon exchange (left) and absolute cross-section of $\pi^{-}\gamma \rightarrow \pi^{-}\pi^{-}\pi^{+}$ derived from it (right), both compared to the respective illustration of the ChPT prediction.}
    \label{fig:xsection}
  \end{center}
\end{figure}

The black points in Fig.~\ref{fig:xsection} (left) show the cross-section of the Primakoff contribution 
 to $\pi^{-}{\rm Pb} \rightarrow \pi^{-}\pi^{-}\pi^{+} {\rm Pb}$, \ie the total
 intensity of the chiral amplitude in the respective mass bins converted by the kaon normalization, 
 with their statistical uncertainties from the PWA.
The triangles picture the ChPT prediction as given by Eq.~\ref{eq:WW-cross-sec} which is 
 integrated in the same mass bins as used in the analysis of the data. 
The full systematic uncertainties include the uncertainty of the luminosity determination from the kaon decays,
 results from different fitting models,
 and estimations for the radiative corrections to be applied \cite{Kaiser_priv}, 
 so that the total uncertainties sum up to about $20\%$.

From this measured cross-section the quasi-real photon density is divided out to get the 
  cross-section of $\pi^{-}\gamma \rightarrow \pi^{-}\pi^{-}\pi^{+}$ presented in Fig.~\ref{fig:xsection} (right).
The ChPT calculation from \cite{KaiserFriedrich08} is depicted here as a continuous line, 
 while the data points are still placed at the center of the respective bin in which they have been obtained, 
 showing good agreement with the theoretical prediction.

\section{Conclusion and Outlook}

The result presented in this paper confirms the ChPT leading order calculation for $\pi^{-}\gamma \rightarrow \pi^{-}\pi^{-}\pi^{+}$ 
with an experimental uncertainty of about 20\%, both in its integrated strength and its shape in the 5 dimensional 
three-body phase space for $m_{3\pi}<0.72\,{\rm GeV}/c^{2}$, for the first time.

For the adjacent region of higher masses, both loop effects and explicit $\rho$ contributions have additionally to be 
 taken into account for the chiral amplitude. 
The need to include isobaric decays with $M=1$ becomes obvious for masses approaching the $a_{2}(1320)$ resonances at latest.
These extensions of the presented analysis are subject of work currently being carried out.

\acknowledgements{%
  This work is supported by the German BMBF, the Maier-Leibnitz-Labor M\"unchen, the DFG Cluster of Excellence Exc153, and CERN-RFBR grant 08-02-91009.
}


%

}  


\end{document}